\documentclass[12pt]{iopart}
\usepackage{fullpage,graphicx,epsf,ulem,float}
\usepackage{iopams}
\usepackage{amsfonts,amssymb}
\usepackage{amscd}
\usepackage{cite}
\usepackage{color}
\usepackage{graphicx}
\usepackage{epsfig}
\usepackage[english]{babel}
\usepackage{amsfonts}
\usepackage{soul}
\begin{document}
\def\be{\begin{equation}}
\def\ee{\end{equation}}
\def\bea{\begin{eqnarray}}
\def\eea{\end{eqnarray}}
\def\fr{\frac}
\def\l{\label}

\newcommand{\gae}{\lower 2pt \hbox{$\, \buildrel {\scriptstyle >}\over {\scriptstyle
\sim}\,$}}
\newcommand{\lae}{\lower 2pt \hbox{$\, \buildrel {\scriptstyle <}\over {\scriptstyle
\sim}\,$}}
\title{Condensate formation in a zero-range process with random site
capacities}
\author{Shamik Gupta$^{1,2,3}$, Mustansir Barma$^4$}
\address{$^1$ Laboratoire de Physique Th\'{e}orique et Mod\`{e}les
Statistiques (CNRS UMR 8626), Universit\'{e} Paris-Sud, Orsay, France}
\address{$^2$ Dipartimento di Fisica e Astronomia,
Universit\`{a} di Firenze, via G. Sansone, 1 50019 Sesto
Fiorentino, Italy}
\address{$^3$ Present address: Max Planck Institute for
Physics of Complex Systems, N\"{o}thnitzer Strasse 38, D-01187 Dresden,
Germany}
\address{$^4$ Department of Theoretical Physics, Tata Institute of
Fundamental Research, Homi Bhabha Road, Mumbai 400005, India}
\ead{shamikg1@gmail.com,barma@theory.tifr.res.in}
\date{\today}
\begin{abstract}
We study the effect of quenched disorder on the zero-range process (ZRP), a system
of interacting particles undergoing biased 
hopping on a one-dimensional periodic lattice, with the disorder
entering through random capacities of sites. In the usual ZRP, sites can
accommodate an arbitrary number of particles, and for a class of hopping
rates and high enough density, the steady state exhibits a condensate
which holds a finite fraction of the total number of particles. The
sites of the disordered
 zero-range process considered here have finite capacities chosen
 randomly from the Pareto
distribution. From the exact steady state measure of the model, we
identify the conditions for condensate formation, in terms of parameters that involve both
interactions (through the hop rates) and randomness (through the
distribution of the site capacities). Our predictions are supported by results obtained from a direct numerical sampling of the steady state and from Monte Carlo
simulations. Our study reveals that for a given realization of
disorder, the condensate can relocate on the subset of sites with
largest capacities. We also study sample-to-sample variation of the critical density
required to observe condensation, and show that the corresponding
distribution obeys scaling, and has a Gaussian or a L\'{e}vy-stable form depending on the values of the relevant parameters. 
\end{abstract}
\maketitle
\tableofcontents

\section{Introduction and model}
\l{sec:intro}
Quenched disorder can strongly affect both static and time-dependent properties
of statistical systems. Of particular 
interest is the case of driven systems in which a dynamics that violates
detailed balance leads the system to a nonequilibrium stationary state
(NESS) that cannot be described within the purview of the
Boltzmann-Gibbs equilibrium statistical mechanics.
In this work, we explore effects of quenched disorder by analyzing a one-dimensional model of a disordered nonequilibrium system,
namely, a disordered zero-range process. The model is a
modification of the well-studied zero-range process (ZRP), a lattice model of interacting
particles evolving in presence of an external drive, with no limit on
the capacity of each site to hold any number of particles
\cite{Spitzer:1970,Evans:2000,Evans:2005,Godreche:2007}. Here, we
study a disordered model introduced in
\cite{Tripathy:1997,Tripathy:1998}, in which the capacity of each site
is a randomly chosen finite number. Knowledge of the exact steady
state of this model allows us to unveil and understand the physical
effects that result from quenched disorder. We are primarily interested
in the possible occurrence of a condensate, which is a quintessential feature
of the ZRP, as discussed below.

On a one-dimensional periodic lattice, the ZRP dynamics involves particles
undergoing stochastic hopping between the lattice sites \cite{Spitzer:1970,Evans:2000,Evans:2005,Godreche:2007}.
For a system of $L$ sites and $N$ indistinguishable unit-mass particles, a unit time step of the dynamics comprises $L$ sequential moves, in each of which a particle hops
out of a random site $i;~i=1,2,\ldots,L$, with occupancy $n_i;~n_i
> 0$, with a specified hop rate $u_i(n_i)$, and moves to site $i + 1$.
The particle density is $\rho \equiv N/L$. The forward-biased hopping of a particle from a site to only its right neighbor
incorporates the effect of an external driving field on the
particles. Evidently, the dynamics conserves the total number of particles in the system.  
While there is no interaction between particles on different sites, that
between particles on the same site may be modeled through the
dependence of the hop rate of a particle from the site on its
occupancy. Remarkably, the NESS measure of configurations in the ZRP can be
found exactly for any choice of hop rates and in any spatial dimension \cite{Spitzer:1970,Evans:2000,Evans:2005,Godreche:2007}.
The homogeneous ZRP is defined by having the same functional form of the hop rate for all sites.

The phenomenon of real-space condensation that can occur in the steady
state of the ZRP involves a finite fraction of particles
accumulating on a single site, thereby forming a macroscopic condensate whose mass increases with increasing density. 
In the case of the homogeneous ZRP, a possible choice of the hop rate is $u(n)= 1+b/n$, where $b>0$ is a finite constant. Such a form of the hop
rate implies an effective attraction between particles on the same site.
For such a choice, it is known that for $b>2$, the model in the NESS exhibits 
a transition to a condensate phase at a critical value of the particle density given by $\rho_c=1/(b-2)$ \cite{Evans:2005}. For densities $\rho < \rho_c$, the system is
in a fluid phase characterized by an occupancy of order unity on every
site and a single-site occupancy distribution $p(n)$ that decays
exponentially for large $n$. At the transition point $\rho=\rho_c$, the
distribution decays asymptotically as a power-law, corresponding to a critical fluid.
Above $\rho_c$, the critical fluid coexists with a macroscopic aggregate
(the ``condensate"), so that in addition to a power-law part, the
distribution $p(n)$ has a sharp peak around $n=(\rho-\rho_c)L$ that
represents the condensate. The ZRP has been invoked to model condensation in a number of contexts,
e.g., clustering of particles in shaken granular systems 
\cite{Torok:2005}, jams in traffic flows \cite{Chowdhury:2000}, wealth
condensation in macroeconomies \cite{Burda:2002}, and other systems.

Driven diffusive systems constitute a class of stochastically evolving
interacting particle systems typified by a spreading of density
fluctuations with a systematic drift in addition to a diffusive motion \cite{Schmittmann:1995,Stinchcombe:2001,Schutz:2001}. At long times, these systems relax to a NESS in which a steady current of particles flows through the system.
The forward-biased ZRP described above is an example of a driven diffusive system, and
can also be mapped to another paradigmatic and extensively
studied model in this class, namely, the
asymmetric simple exclusion process (ASEP). On a one-dimensional
periodic lattice, the ASEP involves indistinguishable hard-core
particles undergoing biased hopping to empty nearest-neighbor sites. The
mapping between the ZRP and the ASEP consists in interpreting sites
(respectively, particles) in the former as particles
(respectively, empty sites) in the latter \cite{Evans:2005}. 

Quenched disorder in driven diffusive systems has been studied over the
years in several types of systems in this class \cite{Barma:2006}. Both particle-wise
disorder, in which different particles have different time-independent
hop rates \cite{Krug:1996,Evans:1996,Juhasz:20051,Juhasz:20052,Masharian:2012},
and space-wise disorder, with time-independent hop rates that are
randomly distributed in space
\cite{Tripathy:1997,Tripathy:1998,Krug:2000,Barma:2002,Jain:2003,Harris:2004,Angel:2004,Enaud:2004}, have been considered. Note that the mapping between the ASEP and the ZRP 
mentioned above transforms particle-wise disorder in
the ASEP into space-wise disorder in the corresponding ZRP
\cite{Krug:1996,Evans:1996}. 
Other studies of quenched disorder in driven systems include a disordered ASEP with
particle non-conservation, in which randomly
chosen sites do not conserve particle number \cite{Evans:2004}, a ZRP on
inhomogeneous networks, in which particles hop between nodes of a network with one node of degree much higher than a typical degree \cite{Waclaw:2007}, a ZRP with quenched disorder in the particle
interaction, implemented through a small perturbation of a generic
class of hop rates \cite{Grosskinsky:2008,Molino:2012}, and a ZRP
involving an interplay between on-site interaction and
 diffusion disorder \cite{Godreche:2012}.

In contrast to the above mentioned studies of quenched disorder, in the
model under consideration here, disorder is assigned to the capacities of sites; the capacity
is a random variable that restricts the number of
particles a site can accommodate. This model was introduced in
\cite{Tripathy:1997,Tripathy:1998} as the disordered drop-push process
(DDPP) to study transport of carriers trapped in local
regions of space, and is a generalization of the uniform drop-push
process \cite{Barma:1993,Schutz:1996}. As pointed out in \cite{Evans:2000}, the drop-push process is actually a special case of the ZRP, with infinite hopping rates out of sites in which the occupancy exceeds the capacity. We thus prefer to refer to the model as
the random capacity zero-range process (RC-ZRP). The fact that
capacities are finite and random has a strong and an essential
influence on the ZRP steady-state dynamics, as we discuss below. 

In this paper, we consider capacities chosen independently for every site from a common distribution with
power-law tails. For a one-dimensional periodic lattice with $L$
sites, every site $i$ has a capacity $l_i$ chosen
independently from the Pareto distribution:
\be
P(l)=\fr{\alpha}{l^{1+\alpha}};~~ \alpha>0, ~{\rm and~} l \in [1,\infty).
\l{eq:Pl}
\ee
Note that for a given realization $\{l_i\}$ of the disorder, the system
can accommodate at most $N_{\rm max}\equiv \sum\limits_{i=1}^L l_i$
particles \footnote{In simulations of the RC-ZRP reported later in the
paper, the capacity of a site is taken as the largest integer not exceeding
a real number drawn from the distribution (\ref{eq:Pl}).}.  

A new feature, namely, a dynamical cascade effect, emerges owing to
sites having restricted capacities in the RC-ZRP
\cite{Tripathy:1997,Tripathy:1998,Barma:1993,Schutz:1996}. Consider a particle hopping out of a random site $i$ that
has occupancy $0<n_i \le l_i$ with hop rate $u_i(n_i|l_i)$, and moving to
site $i + 1$. If the site $i+1$ is already full, a particle from
this site gets pushed further right, and so on, leading to a sequence of
adjacent-site hops that continues until a particle hops into a site
($i + m$, say) that was not fully occupied earlier (i.e., $n_{i+m}
<l_{i+m}$). Note that a unit time step corresponds to $L$ updates at randomly chosen sites, where each update may involve several particle hops out of fully filled sites. Thus, restricted capacities lead to
a cascade of particle transfers through filled sites, explaining the
nomenclature ``drop-push process" used in \cite{Barma:1993,Schutz:1996}. Figure \ref{fig:dzrp}
shows a schematic view of the RC-ZRP. 

\begin{figure}
\centering
\includegraphics[width=100mm]{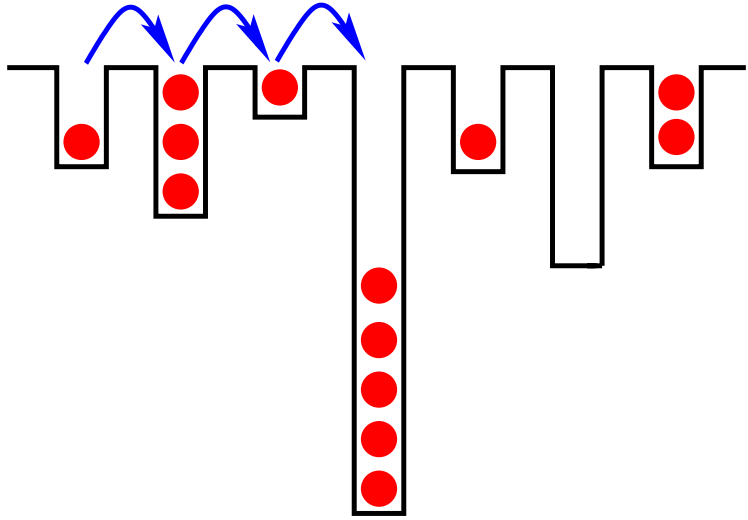}
\caption{Schematic diagram of the random capacity zero-range process, showing a typical configuration and possible
particle hops.}
\l{fig:dzrp}
\end{figure}

In this paper, we ask:
Does restricting the ZRP site capacity, as in the RC-ZRP, still allow
for the formation of a condensate, and if so, under what conditions? Let
us consider a form for the hop rate $u_i(n|l_i)$ that promotes condensate formation in the homogeneous ZRP, namely,
$u_i(n>0)=1+b/n~\forall~i$, with $b>0$. The hop rate is taken to have this form for occupancies in the range $0
< n \le l_i$, and to be infinite for occupancies larger than $l_i$
\cite{Evans:2000}, ensuring an immediate movement of a particle from a
filled site to one that is not full:
\be
u_i(n|l_i)=\left\{\begin{array}{cc}
          0 & \mbox{for~} n=0, \\
          1+\fr{b}{n} & \mbox{for~} 0 < n \le l_i, \\
          \infty    & \mbox{for~} n > l_i.
          \end{array}
          \right.         
\l{eq:u}          
\ee
Note that in contrast to earlier studies of the model in \cite{Tripathy:1997,Tripathy:1998} which considered general hop
rates, the choice (\ref{eq:u}) has the possibility of supporting condensate formation.
In this paper, we demonstrate on the basis of exact
analytical and simulation results that an interplay of the capacity distribution (\ref{eq:Pl}) with the hop rate
(\ref{eq:u}) can indeed lead to condensate formation, and derive the
conditions for this to happen. It should be noted that the steady state equal-time
properties reported in this paper hold not just for the considered case of biased hopping of particles from a site to its right
neighbor, but also for unbiased hopping to the left and to the right
neighboring site. This is because, as we discuss in Section
\ref{sec:steady-state}, the steady state measure of the RC-ZRP is the
same in the two cases. However, unequal-time properties in the steady state, for instance,
relocation dynamics of the condensate, will be different for biased and unbiased hopping.

The ZRP in which sites have bounded capacities was addressed recently
in \cite{Ryabov:2014}. Unlike our model, the
capacity was taken to be the same for all sites, and quenched disorder
was introduced through site-dependent and particle-dependent hop-rates,
leading to dynamical blocking that causes slow relaxation to steady state.

The layout of the paper is as follows. In Section \ref{sec:steady-state}, we discuss the steady
state of the RC-ZRP, based on which we derive in
Section \ref{sec:condensation-condition} the conditions to obtain condensation in the
model. In Section \ref{sec:simulations}, we confirm our predictions by a direct numerical sampling of the steady state and by performing Monte Carlo
simulations of the steady-state dynamics. We also discuss the
relocation dynamics of the condensate. The paper ends with conclusions in
Section \ref{sec:conclusions}, while the
Appendix summarizes some relevant features associated with the capacity
distribution.

\section{Stationary state}
\l{sec:steady-state}
The RC-ZRP relaxes at long times to a
current-carrying nonequilibrium stationary state.
Using the condition of pairwise balance \cite{Barma:1993}, as in \cite{Tripathy:1997,Tripathy:1998}, the steady state measure of
configurations may be found. For a
given realization $\{l_i\}$ of the disorder and a given total number of
particles $N \le N_{\rm max}$, it has a factorized form \footnote{The measure (\ref{eq:pni})
holds in any spatial dimension, for any choice of the hop rate, and for
any rules of particle transfer, either biased or unbiased, between
sites. In case of unbiased transfer, a case not addressed here, detailed
balance holds, and the steady state is an equilibrium state.} :
\be
{\cal
P}(\{n_{i}\}|\{l_i\},N)\propto \prod\limits_{i=1}^{L}f_{i}(n_{i}|l_i)\delta\left(\sum\limits_{i=1}^{L}n_{i},N\right),
\l{eq:pni}
\ee
where $\delta(m,n)$ is the Kronecker Delta
function, while the single-site factors $f_{i}(n|l_i)$ equal unity for $n=0$ and are
given for $n>0$ by
\bea
f_{i}(n|l_i)&\equiv&\left[\prod\limits_{m=1}^{n}u_i(m|l_i)\right]^{-1}\l{eq:fin-general}\\
&=&\left\{\begin{array}{cc}
          \fr{\Gamma(b+n+1)}{\Gamma(n+1)\Gamma(b+1)} & \mbox{for~} 0 < n \le l_i, \\
          0    & \mbox{for~} n > l_i,
          \end{array}
          \right.  
\l{eq:fin}
\eea
where $\Gamma(x)$ is the Gamma function. Here, in arriving at the second
equation, we have used Eq. (\ref{eq:u}).

Equation (\ref{eq:pni}) is the measure of configurations within the
canonical ensemble. In the thermodynamic limit $N \to \infty, L \to
\infty$, keeping the overall particle density $\rho \equiv N/L$ fixed,
we use an equivalent grand canonical ensemble
description of the steady state. In such a description, the total number
of particles is allowed to fluctuate, and a fugacity $z$ fixes the average number of particles to equal $N$. The
steady state measure of configurations within the grand canonical
ensemble is given by
\be
{\rm Prob}(\{n_{i}\}|\{l_i\}) \propto \prod_{i=1}^L p_i(n_i|l_i),
\l{eq:GC-full}
\ee
where $p_i(n|l_i)$ is the single-site occupancy distribution, namely, the
probability for the $i$th site to
have $0 \le n \le l_i$ particles:
\be
p_i(n|l_i)\equiv\fr{z^n f_i(n|l_i)}{F_i(z|l_i)},
\l{eq:GC-singlesite}
\ee
with $F_i(z|l_i)$ ensuring normalization of $p_i(n|l_i)$:
\be
F_i(z|l_i) \equiv 1+\sum_{n=1}^{l_i}z^nf_i(n|l_i).
\ee
Here, the fugacity $z$ satisfies $\sum_{i=1}^L \overline{n_i}=N$, where
$\overline{n_i}=d\ln F_i(z|l_i)/d\ln z$ is the average occupancy at the $i$th site, with
the average taken with respect to the single-site probability
(\ref{eq:GC-singlesite}). We finally get 
\be
\fr{1}{L}\sum_{i=1}^L \fr{zF_i'(z|l_i)}{F_i(z|l_i)}=\rho,
\l{eq:z}
\ee
where prime denotes differentiation with respect to $z$.

As discussed in the Appendix, the maximum number of particles $N_{\rm
max}$ that can be accommodated in the system scales with $L$ as $N_{\rm
max} \sim L^{1/\alpha}$ for $\alpha<1$, and as $N_{\rm max}=aL$ for
$\alpha>1$, where $a \equiv \int dl~lP(l)$ is finite. It then follows that in the latter
case, the density $\rho$ in Eq. (\ref{eq:z}) has a maximum allowed
finite value
equal to $a$, while for $\alpha<1$, $\rho$ diverging with $L$ as
$L^{1/\alpha-1}$ can be arbitrarily large. For a given value of $z$, evaluating numerically the left hand side of Eq.
(\ref{eq:z}) for a given realization of disorder, and then averaging with respect to disorder, we show in 
Fig. \ref{fig:z-rho} the disorder-averaged $\rho$, denoted by $\langle
\rho \rangle$, as a function of $z$ for three different system sizes \footnote{Here and in the rest of the paper, angular brackets will be
used to denote averaging with respect to disorder realizations.}. 

\begin{figure}[H]
\centering
\includegraphics[width=100mm]{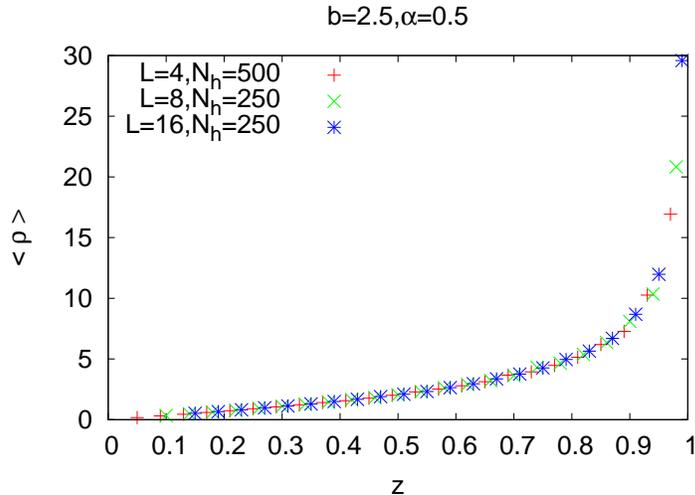}
\caption{Disorder-averaged density $\langle \rho \rangle$ versus
fugacity $z$, computed using Eq. (\ref{eq:z}) for $b=2.5$ and
$\alpha=0.5$. The system sizes are marked in the figure. Here,
$N_h$ is the number of disorder realizations over which the data have been averaged.}
\l{fig:z-rho}
\end{figure}

The occupancy distribution for the
full system is defined as the probability that a
randomly chosen site has $n$ particles. This is possible only if the
site capacity is equal to or larger than $n$, so that the distribution has
the form
\be
\widetilde{p}(n) \propto z^n f(n) \int_n^\infty dl~P(l).
\l{eq:GC-singlerandomsite}
\ee

For large $n \gg 1$, Eq. (\ref{eq:fin}) combined with the approximation $\Gamma(b+n+1)/\Gamma(n+1)\approx
n^{b}+O(n^{b-1})$ yields the following asymptotic behavior:
\bea
&&p_i(n|l_i) \propto \fr{\exp(-n/n^\star)}{n^b}, \l{eq:GC-singlesite-pn} \\
&&\widetilde{p}(n) \propto \fr{\exp(-n/n^\star)}{n^{b+\alpha}},
\l{eq:GC-singlesite-ptilden}
\eea
where the characteristic occupancy $n^\star$ is given by
\be
n^\star\equiv -1/\ln(z).
\l{eq:nstar}
\ee

The RC-ZRP in the NESS supports a steady current of particles through
the system. Within the grand canonical ensemble, an exact expression for the average steady-state current for a given
realization of disorder was derived in \cite{Tripathy:1998}.
We briefly summarize the derivation here. From the dynamics, it is
evident that all hops contributing to the current $J_{i-1,i}$ across the
bond $(i-1,i)$ for which site $i$ is completely full also contribute to
$J_{i,i+1}$, so that one has the recursion 
\be
J_{i,i+1}=p_i(l_i|l_i)J_{i-1,i}+\sum_{n_i=1}^{l_i}u_i(n_i|l_i)p_i(n_i|l_i),
\l{eq:J-3}
\ee
where the second term on the right hand side is due to hops
originating from site $i$. Now, steady state implies that all the bond currents are equal,
$J_{i,i+1}=J_{i-1,i}=\ldots=J_0$, so that the above equation yields
\be
J_0={{\sum_{n_i=1}^{l_i}u_i(n_i|l_i)p_i(n_i|l_i)}
             \over{1-p_i(l_i|l_i)}}=z,
\l{eq:J-4}
\ee
where we have used Eqs. (\ref{eq:fin-general}) and
(\ref{eq:GC-singlesite}). The steady-state current depends through the fugacity $z$ on
the overall particle density $\rho$ and the number of sites of
different capacities in a given realization of disorder (see Eq.
(\ref{eq:z})), thus becoming a function of $b$ and
$\alpha$. Note that the expression (\ref{eq:J-4}) for the average steady-state current
in terms of the fugacity $z$ is the same as for the usual ZRP \cite{Evans:2005}.

In the following section, we address the issue of condensate formation in the RC-ZRP.

\section{Condensate formation}
\l{sec:condensation-condition}
In this section, we turn to the conditions for condensate formation in
the RC-ZRP. We also study the distribution
of the critical density $\rho_c$ to observe condensation, and its scaling as a function of $L$.
We begin by summarizing the main questions and the results before getting to the
details of the derivation. 

It is useful to first recall the known scenario of condensation
in the customary homogeneous ZRP \cite{Evans:2005}. Since there is no restriction on site
capacities, there is no difficulty in accommodating $O(L)$ particles on
any one site. An essential
requirement for condensation is
the existence of a finite critical value of the average site occupancy
in the limit the fugacity $z$ attains its maximum possible value $z_{\rm
max}$. Below this critical value of the density, all sites have the same average
occupancy equal to the overall density of particles in the system. Above the critical value, the
average occupancy of all but one site has the
critical value; the excess particles that form a finite fraction of the
total number of particles are accommodated on a single randomly-chosen
site. 

In this backdrop, it is a priori not apparent whether and when such a
scenario of condensation holds in the RC-ZRP in which sites have
restricted capacities. To address the issue, we argue as follows.
\begin{enumerate}
\item A necessary condition for condensate formation is that at least
one site be able to accommodate $O(L)$ particles. In view of sites
having restricted capacities in the RC-ZRP, the candidate for a
site that can accommodate a macroscopic number of particles is the one
with the largest capacity. Then, if the largest capacity $l_{\rm max} \equiv {\rm Max}[l_1,l_2,\ldots,l_L]$ has the
scaling $l_{\rm max} \sim L^\theta$, we need $\theta$ to be larger
than unity for condensate formation. 
\item Additionally, we require that the average site occupancy has a
finite value, denoted by $\rho_c$, as $z \to z_{\rm
max}$. 
\item When conditions (i) and (ii) are fulfilled, condensate formation is possible at
high enough density $\rho > \rho_c$. The critical density $\rho_c$
depends on the realization of disorder, and the question arises as to
how the form of the disorder-induced distribution ${\rm Prob}(\rho_c)$ of
$\rho_c$ depends on the relevant parameters.
\end{enumerate}

Whether conditions (i) and (ii) above would hold depends on parameters that
characterize the probability distribution of capacities (Eq.
(\ref{eq:Pl})) and the form of the hop rate (Eq. (\ref{eq:u})), namely,
the exponents $\alpha$ and $b$. The results are as follows:
\begin{enumerate}
\item The exponent $\theta$ is given by $1/\alpha$, so the site with the
largest capacity can accommodate $O(L)$ particles provided that $\alpha<1$.
\item The average site occupancy remains finite as $z
\to z_{\rm max}$ so long as one has $b+\alpha>2$.

Combining the last two points, we thus arrive at the following conditions
for condensate formation in the RC-ZRP:
\be
b+\alpha >2, ~~~~{\rm~and~} \alpha<1 ~~~~~({\rm Conditions~to~obtain~condensation}).
\l{eq:cond-condensation}
\ee
\item The distribution ${\rm Prob}(\rho_c)$ is a Gaussian
for $b > (4-\alpha)/2$, while it is a L\'{e}vy-stable distribution for $1
< b < (4-\alpha)/2$.
\end{enumerate}
These results are summarized in Fig. \ref{fig:ph-diag}, which shows the regime for
condensate formation in the $\alpha-b$ plane, and also the forms of
${\rm Prob}(\rho_c)$ in different regions.

The issue of sample-dependence of phase transitions was studied numerically in \cite{Enaud:2004}
 for the one-dimensional ASEP with open boundaries and quenched-disordered hopping rates. In the present work, we are able to determine the analytic forms for the distribution of the critical density because of the product form of the steady state measure in the RC-ZRP with periodic boundary conditions.

In the limit $\alpha \to 0$, the capacities become infinitely
large, and the RC-ZRP dynamics becomes similar to the dynamics of the
homogeneous ZRP. In this limit, the condition to observe condensation becomes the 
requirement $b>2$, a result known for the homogeneous ZRP \cite{Evans:2005}. 

We now proceed to a derivation and a more detailed discussion of our results.

\begin{figure}[H]
\centering
\includegraphics[width=100mm]{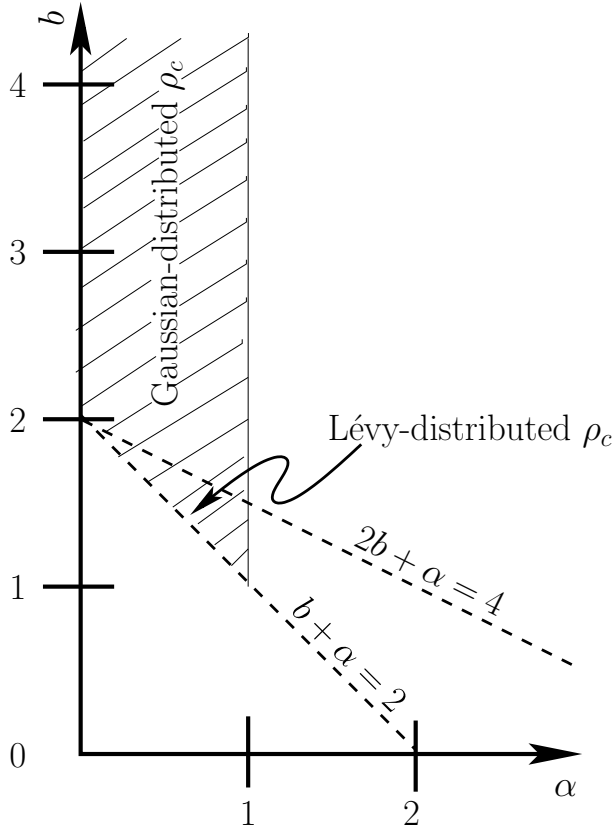}
\caption{In the $\alpha-b$ plane, the dashed regions
are those for which the system exhibits a condensate; the corresponding
forms of the $\rho_c$ distribution are also indicated.}
\l{fig:ph-diag}
\end{figure}

\subsection{The largest capacity $l_{\rm max}$}
\l{sec:max-site}
In order to accommodate the condensate, it is necessary that $l_{\rm
max}$ grows sufficiently rapidly with $L$, namely, as $L^\theta$, with $\theta > 1$.
For the Pareto distribution of Eq. (\ref{eq:Pl}), it is known that
$l_{\rm max}$ scales as $L^{1/\alpha}$, see the Appendix. Thus,
condensate formation is possible provided that $\alpha \le 1$. 

As discussed in the Appendix, not just the site with capacity $l_{\rm
max}$, but in fact several sites have capacities of order $L^{1/\alpha}$. This
feature allows the condensate to form and relocate in time on other
sites that belong to this subset. The numerical studies reported in
Section \ref{sec:simulations} bear this out.

\subsection{Average site occupancy}
\l{sec:site-occupancy-behavior}
Let us now investigate the behavior of the average site
occupancy as $z \to z_{\rm max}$, in the regime $\alpha<1$ as required to accommodate
a putative condensate (Section \ref{sec:max-site}). The quantity $z_{\rm
max}$ is obtained by requiring
the convergence of the series $F_{i_1}(z)\equiv 1+\sum\limits_{n=1}^{l_{\rm
max}}z^{n}f_{i_1}(n)$ in the limit $L\to \infty$, where $i_1$ is the label for the site with the
largest capacity. We obtain $z_{\rm max}=1$ as the radius of
convergence of the series. The average site occupancy being a monotonically
increasing function of $z$ has a maximum allowed value in the limit $z
\to z_{\rm max}$, given by 
\bea
\rho^\star(\{l_i\}) &\equiv& \lim_{z \to z_{\rm max}}\fr{1}{L}\sum_{i=1}^L
\fr{zF_i'(z|l_i)}{F_i(z|l_i)}\nonumber \\
&=&\fr{1}{L}\sum\limits_{i=1}^L G(l_i);~~~~G(l_i)
\equiv \fr{\sum_{n=1}^{l_i} nf_i(n|l_i)}{1+\sum_{n=1}^{l_i}f_i(n|l_i)},
\l{eq:rhostar}
\eea
which defines a characteristic density $\rho^\star$ for every disorder
realization. Evidently, the density $\rho^\star(\{l_i\})$ is given by a sum of $L$ i.i.d. random
variables $G(l_i);~i=1,2,\ldots,L$. A finite value of $\rho^\star$
implies condensate formation for $\rho > \rho^\star$, so that
$\rho^\star$ coincides with the critical density $\rho_c$ to obtain
condensation.
From Eq. (\ref{eq:rhostar}), we get the corresponding disorder-averaged
characteristic density as
\be
\langle \rho^\star \rangle \equiv \int_{1}^{\infty}dl~P(l)G(l).
\l{eq:rhostar-sa}
\ee

\begin{figure}
\centering
\includegraphics[width=100mm]{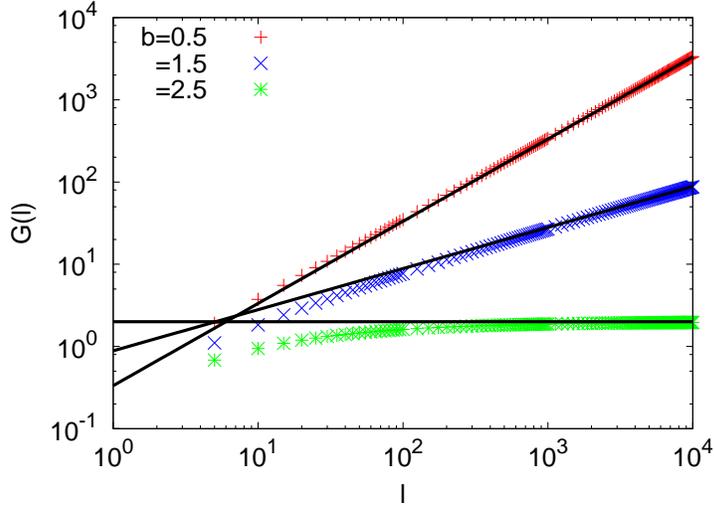}
\caption{$G(l)$ as a function of $l$ for three
representative values of $b$, obtained by evaluating numerically Eq.
(\ref{eq:Gl-exact}). The black lines stand for the
asymptotic behavior, Eq.~(\ref{eq:Gl-asymp}).}
\l{fig:Gl-exact-ap}
\end{figure}

The behavior of $\rho^\star$ is governed by two parameters, namely, the
exponent $\alpha$ that characterizes the probability distribution of the
capacities, and the exponent $b$ that characterizes the hop rate. Let us
ask for the condition on the allowed range of values
of $b$ and $\alpha$ that leads to a
finite $\langle \rho^\star \rangle$. Using Eq. (\ref{eq:fin}), the function $G(l)$ in
Eq.~(\ref{eq:rhostar}) can be expressed as
\be
G(l)=\fr{1}{b-2}+\fr{(b-1)(l+1)}{(b-2)\Big(1-\Gamma(b+l+1)/[\Gamma(b)\Gamma(l+2)]\Big)},
\l{eq:Gl-exact}
\ee
implying that $G(l)$ diverges for particular values $b=1,2$. In the
asymptotic regime $l \gg 1$, using
$\Gamma(b+l+1)/\Gamma(l+2)\approx l^{b-1}+O(l^{b-2})$, we get
\be
G(l \gg 1)\approx\left\{\begin{array}{cc}
          \fr{(1-b)l}{2-b} & ~~~~~\mbox{for~} 0 < b < 1, \\
          \fr{(b-1)\Gamma(b)l^{2-b}}{2-b} & ~~~~~\mbox{for~} 1 < b < 2, \\
          \fr{1}{b-2} & \mbox{for~} b > 2. \\
          \end{array}
          \right.
\l{eq:Gl-asymp}
\ee
Figure \ref{fig:Gl-exact-ap} illustrates that the asymptotic form, Eq.
(\ref{eq:Gl-asymp}), is a good approximation to the exact expression,
Eq.~(\ref{eq:Gl-exact}). Now, Eq. (\ref{eq:rhostar-sa}) gives
\bea
\langle \rho^\star \rangle\approx r(l^\star)+\int_{l^\star}^{\infty}dl~P(l)G(l),
\l{eq:rhostar-sa1}
\eea
where $l^\star$ is chosen such that for $l>l^\star$, the function $G(l)$ is well
approximated by its asymptotic behavior,
Eq.~(\ref{eq:Gl-asymp}), for large
$l$. The value of $l^\star$ depends on $b$; for example, from Fig.
\ref{fig:Gl-exact-ap}, one may choose $l^\star=100$ for $b=0.5$ and
$l^\star=1000$
for $b=1.5$. In Eq.~(\ref{eq:rhostar-sa1}), the 
finite constant $r(l^\star)$ is the value of the integral
$\int_1^{l^\star}dl~P(l)G(l)$. By analyzing the integral in
Eq.~(\ref{eq:rhostar-sa1}), one then concludes that requiring $\langle
\rho^\star \rangle$ to be finite leads to the following conditions:
\bea
&&\alpha>1 ~~~~~~~~~\mbox{~for~} 0 < b < 1, \nonumber \\
&&b+\alpha>2 ~~~~\mbox{~for~} 1 < b < 2, \\
&&\alpha>0 ~~~~~~~~~\mbox{~for~} b > 2. \nonumber
\l{eq:b-alpha-conditions-0}
\eea
The above conditions may be combined into the single condition
\be
b+\alpha>2 ~~~~\mbox{~for~} \langle \rho^\star \rangle \mbox{~to~be~finite}.
\l{eq:b-alpha-conditions}
\ee
This is to be contrasted with the condition $b>2$ in the homogeneous ZRP
for the average site occupancy to be finite as $z \to z_{\rm max}$.

At the critical point, when $z \to z_{\rm max}$, the characteristic
occupancy $n^\star$ in Eq. (\ref{eq:nstar}) diverges, and we find from Eqs.
(\ref{eq:GC-singlesite-pn}) and (\ref{eq:GC-singlesite-ptilden}) the occupancy
distributions for $n \gg 1$ to obey
\bea
&&\lim_{z \to z_{\rm max}} p_i(n|l_i) \propto
\fr{1}{n^b},\l{eq:GC-singlesite-critical-pn} \\
&&\lim_{z \to z_{\rm max}} \widetilde{p}(n) \propto
\fr{1}{n^{b+\alpha}}. \l{eq:GC-singlesite-critical-ptilden}
\eea

\subsection{Distribution of the characteristic density $\rho^\star$}
\l{sec:rhostar-distribution}
For given values of $L$, $\alpha$, and $b$, one may obtain the density $\rho^\star(\{l_i\})$ for different
disorder realizations $\{l_i\}$ by using Eq.
(\ref{eq:rhostar}). Let us denote the corresponding
distribution as ${\rm Prob}(\rho^\star)$. One may deduce the form of ${\rm
Prob}(\rho^\star)$ by invoking the well-known theory of stable
distributions, which concerns the sum $S_L\equiv
\sum_{i=1}^L X_i$ of a number $L$ of mutually independent random
variables $\{X_i;~i=1,2,\ldots,L\}$ having a common distribution
\cite{Feller1,Feller2,Gnedenko:1955}. When this common
distribution has a power-law tail decaying as $|X|^{-1-\alpha}$, then,
for $0 < \alpha <2$, the limiting distribution for $S_L$ as $L \to \infty$
converges in form to a stable distribution that has a tail decaying as  
$|S_L|^{-1-\alpha}$. For $\alpha \ge 2$, on the other hand, the distribution converges to a Gaussian distribution.

From the large-$l$ behavior of $G(l)$ in Eq. (\ref{eq:Gl-asymp}), we
deduce the following tail behavior of ${\rm Prob}(G)$:
\be
{\rm Prob}(G)\sim\left\{\begin{array}{cc}
          G^{-1-\alpha} & ~~~~~\mbox{for~} 0 < b < 1, \\
          G^{-1-\nu};~\nu \equiv \alpha/(2-b) & ~~~~~\mbox{for~} 1 < b < 2, \\
          \Theta\Big(\fr{1}{b-2}-G\Big) & \mbox{for~} b > 2, \\
          \end{array}
          \right.
\l{eq:Gl-tail}
\ee
where $\Theta(x)$ is the unit step function, equal to unity for
$x>0$ and zero otherwise. The above predictions for the tails may be
checked against numerically computed ${\rm Prob}(G)$ using Eq. (\ref{eq:Gl-exact}). We show in Fig.
\ref{fig:Gl-prob} a comparison between numerical
results and our predictions for three representative values of $b$ for $\alpha=0.375$. 
\begin{figure}
\centering
\includegraphics[width=186mm]{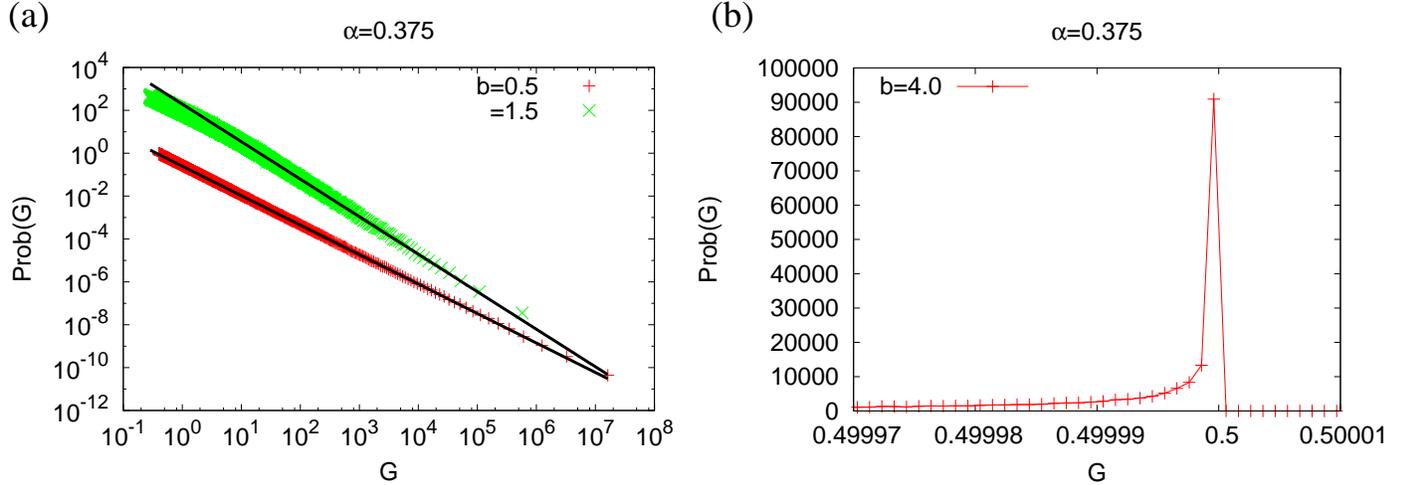}
\caption{Illustrating the validity of the behavior (\ref{eq:Gl-tail}) of
the probability distribution ${\rm Prob}(G)$ for three representative
values of $b$ at $\alpha=0.375$. The values of $G$ are computed using
Eq. (\ref{eq:Gl-exact}). The data for $b=1.5$ have been scaled up
by a factor of $300$ for convenience of display. The black lines denote
analytical predictions, namely, (i) for $b<2$, a power-law decay with exponent $(1+\alpha)$ for $0<b<1$, and exponent $(1+\nu)$ for $1<b<2$, and (ii) for
$b>2$, a unit step function.}
\l{fig:Gl-prob}
\end{figure}

\begin{figure}
\centering
\includegraphics[width=186mm]{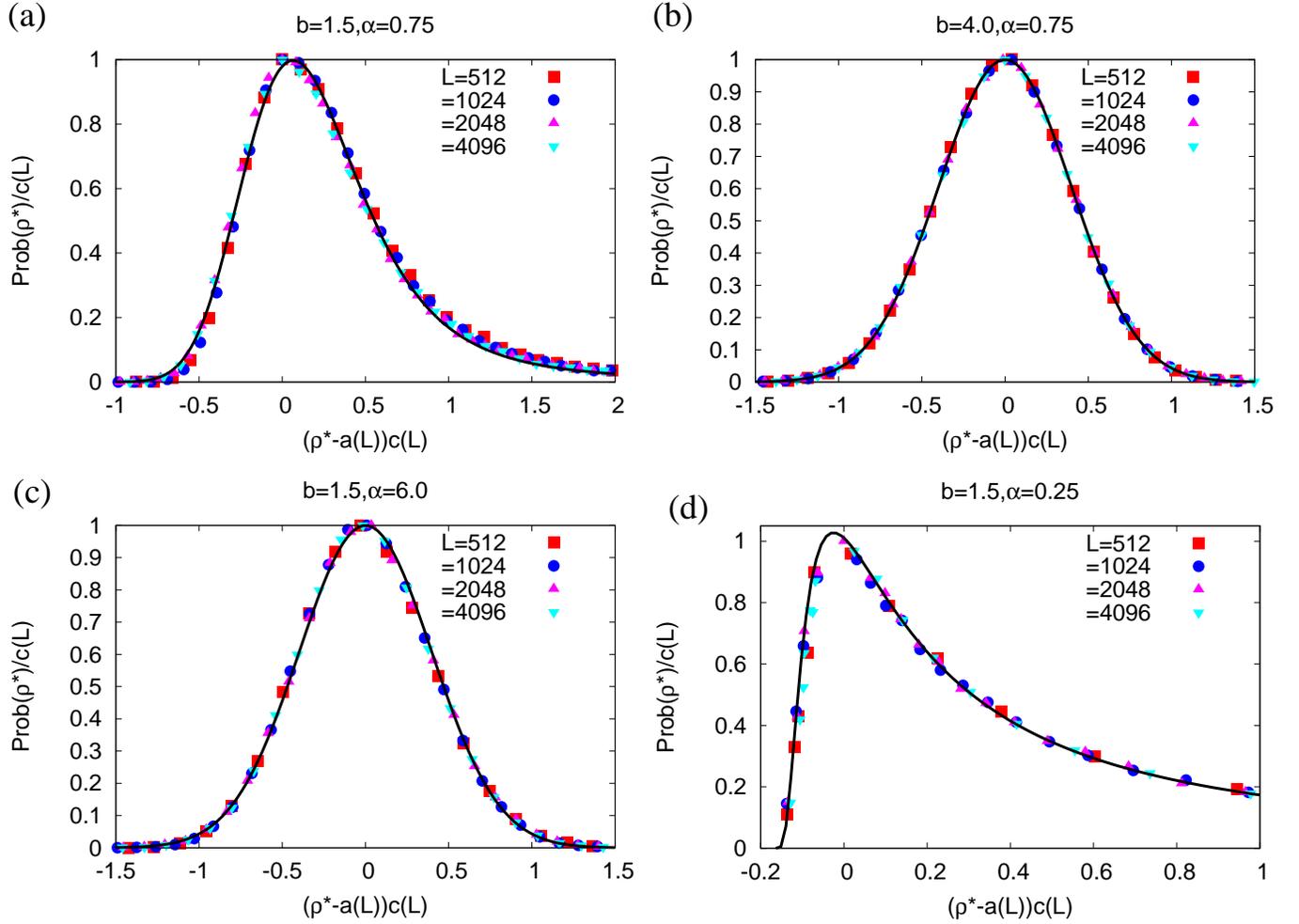}
\caption{Scaling plots for the probability distribution ${\rm
Prob}(\rho^\star)$ for four representative values of the set
$(b,\alpha)$, with $\rho^\star$ computed using Eq. (\ref{eq:rhostar}). Here, the scaling factors are $c(L)={\rm Prob}^{\rm
max}(\rho^\star)$, the maximum value of the distribution, and $a(L)$ the
corresponding value of $\rho^\star$. (a) and (b) stand
for cases where the system supports condensate formation (see condition
(\ref{eq:cond-condensation})), and $\rho^\star$ coincides
with the critical density $\rho_c$. (c) and (d) denote
cases where the system does not support condensation. The black lines denote
analytical predictions, namely, (i) for
$b=1.5,\alpha=0.75$ (giving $\nu=1.5$), the inverse Fourier transform of
the function $\exp[-|c_0k|^\nu(1- i\beta ~{\rm sgn}(k) \Phi)]$, with $\Phi \equiv \tan(\pi
\nu/2)$, $\beta=0.995$, $c_0=0.28$, which has been shifted by an amount
equal to $0.39$ to fit the data, (ii) for $b=4.0,\alpha=0.75$ and
$b=1.5,\alpha=6.0$, a Gaussian with zero mean and variance equal to
$1/(2\pi)$, and (iii) for $b=1.5,\alpha=0.25$ (giving $\nu=0.5$), the so-called L\'{e}vy distribution
$f(x)=\sqrt{c_0/(2\pi)}\exp[-c_0/(2(x-\mu))]/(x-\mu)^{3/2}$; $c_0=0.45$ and
$\mu=-0.175$.} 
\l{fig:rhostar-prob}
\end{figure}

Invoking the results on stable distributions discussed above,
we may deduce the behavior of ${\rm Prob}(\rho^\star)$ for different
range of values of $b$ by using Eqs. (\ref{eq:rhostar}) and (\ref{eq:Gl-tail}). 
\begin{enumerate}
\item $0<b<1$: Here, ${\rm Prob}(\rho^\star)$ is a L\'{e}vy-stable
distribution with a tail decaying as a power law with exponent $(1+\alpha)$ for values of $\alpha$ in
the range $0 < \alpha <2$ and is a Gaussian distribution for $\alpha \ge
2$. The mean is finite for $\alpha >1$.
\item $1 < b < 2$: In this case, ${\rm Prob}(\rho^\star)$ is a L\'{e}vy-stable
distribution with a tail decaying as a power law with exponent $(1+\nu)$ for values of $\nu$ in
the range $0 < \nu <2$, that is, provided $\alpha < 4-2b$. On the other hand, for $\alpha \ge 4-2b$,
the distribution is a Gaussian. The mean $\rho^\star$ is finite for
$b+\alpha>2$. 
\item $b>2$: In this regime, ${\rm Prob}(G)$ has a finite variance,
implying that ${\rm Prob}(\rho^\star)$ is a Gaussian; the mean is of course
finite.
\end{enumerate}
Thus, the condition $b+\alpha>2$ ensures a finite value of the mean $\rho^\star$, a
condition we derived earlier, see Eq. (\ref{eq:b-alpha-conditions}), based on an analysis of the
disorder-average $\langle \rho^\star \rangle$ defined in
Eq. (\ref{eq:rhostar-sa}).

As a function of the system size $L$, the probability distribution ${\rm
Prob}(\rho^\star)$ has the scaling form (see Fig.
\ref{fig:rhostar-prob}):
\be
{\rm Prob}(\rho^\star) \sim c(L){\mathcal
G}\Big[\Big(\rho^\star-a(L)\Big)c(L)\Big],
\l{eq:rhostar-scaling-form}
\ee
where $c(L)$ is the maximum value of the
distribution, while $a(L)$ is the corresponding value of $\rho^\star$. The scaling
function ${\mathcal G}(x)$ has either (a) a Gaussian form, in which case $a(L)$ is
independent of $L$, and $c(L) \sim \sqrt{L}$, or, (b) a L\'{e}vy-stable form,
in which case we have 
\begin{enumerate}
\item for $0<b<1$: $a(L) \sim L^{1/\alpha-1}$ and $c(L) \sim
L^{1-1/\alpha}$ for $0<\alpha <1$, and $a(L)$ independent of $L$ and
$c(L) \sim L^{1-1/\alpha}$ for $\alpha \ge 2$,
\item for $1<b<2$: $a(L) \sim L^{1/\nu-1}$ and $c(L) \sim L^{1-1/\nu}$
for $0<\nu <1$, and $a(L)$ independent of $L$ and $c(L) \sim L^{1-1/\nu}$
for $\nu \ge 
2$.
\end{enumerate}

\section{Numerical studies}
\l{sec:simulations}
In this section, we check our predictions on the existence of 
a condensate in the RC-ZRP by reporting on results obtained by a direct numerical sampling of its canonical steady state measure for a given realization
of the disorder. When relevant, we also perform Monte Carlo (MC) 
simulations of the RC-ZRP dynamics while starting from the steady state.

The steady state is generated according to the following algorithm.
For a given system size $L$ and disorder realization $\{l_i\}$, a
configuration of the system corresponding to a total $N$ particles is generated by occupying $L-1$ sites independently with
$n_i$ ($0 \le n_i \le l_i$) particles with respective weights
$f_i(n_i|l_i)$; here, $i=1,2,\ldots,L-1$. The deficit number of
particles, $n_{\rm d}\equiv N-\sum_{i=1}^{L-1}n_i$, when positive, is
accommodated on the
last remaining site $i=L$
with the weight $f_L(n_d|l_L)$. If the deficit is negative, the
configuration is rejected, and the process is repeated all over again.
A configuration so generated is run for a typical ``equilibration" time
of order $L$ before performing any analysis of the data. 

\begin{figure}[H]
\centering
\includegraphics[width=186mm]{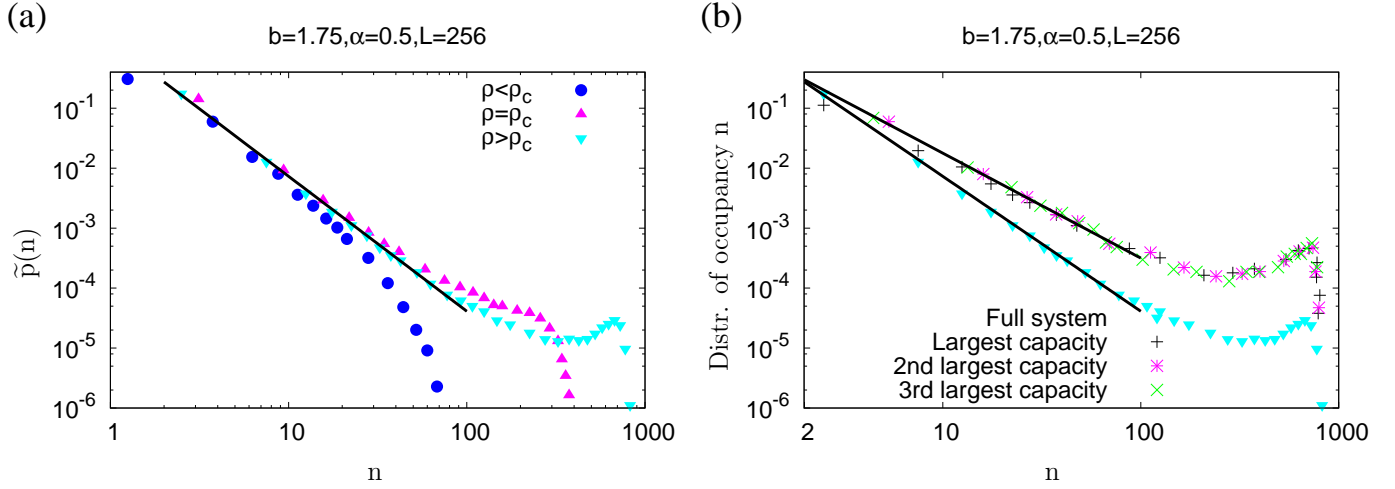}
\caption{For a fixed disorder realization, (a) shows the
occupancy distribution for the full system in the steady state, with densities smaller than, equal to, and larger than the critical density
$\rho_c$; the density values are $1.0$, $2.28$ and $4.0$,
respectively. The system size is $L=256$, and we have taken
${\alpha}=0.5,b=1.75$. The data are obtained by sampling the canonical
steady state measure using the algorithm detailed in the text. Below $\rho_c$, the distribution shows an
exponential decay at 
large $n$, while at $\rho_c$, one has a power-law decay at large $n$
with exponent $(b+\alpha)$, see
Eqs. (\ref{eq:GC-singlesite-ptilden}) and
(\ref{eq:GC-singlesite-critical-ptilden}). For $\rho > \rho_c$, in addition to the
power-law behavior, a bump indicating the presence of a condensate
appears. The black line stands for the power-law behavior $\sim
n^{-(b+\alpha)}$. (b) shows for $\rho > \rho_c$ the occupancy distribution in the steady state for the full
system as well as for sites with the largest, the second largest and the
third largest capacity. Besides a bump at large $n$ that implies the
presence of a condensate, one has a power-law decay at small $n$, with
exponent $(b+\alpha)$ for the full
system, and with exponent $b$ for individual sites, see Eq.
(\ref{eq:GC-singlesite-critical-pn}).} 
\l{fig:Pn-alphalt1}
\end{figure}

\begin{figure}[H]
\centering
\includegraphics[width=100mm]{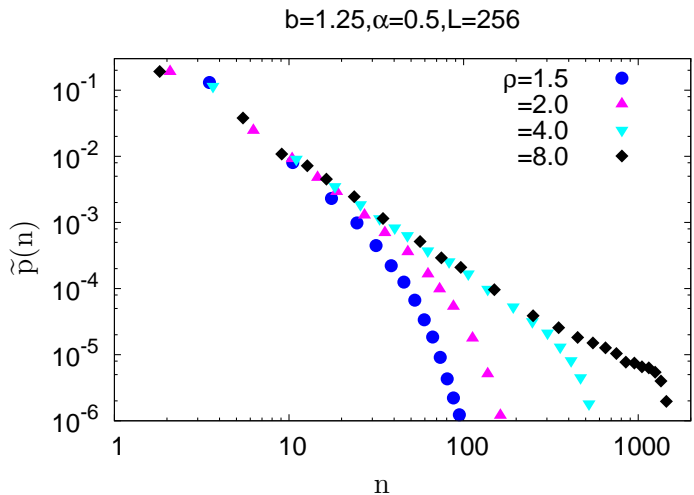}
\caption{For the same disorder realization, system size and $\alpha$ as in Fig.
\ref{fig:Pn-alphalt1}, but for a value of $b$ that does not
satisfy the conditions (\ref{eq:cond-condensation}) to observe condensation, the figure shows the
occupancy distribution for the full system in the steady state at
several densities. The data are obtained by sampling the canonical
steady state measure using the algorithm detailed in the text. In contrast to Fig. \ref{fig:Pn-alphalt1}(a), the distribution at all densities has an exponential decay at
large $n$, and there is no extra peak appearing at high densities that
corresponds to a condensate. Here, we have taken $b=1.25$.} 
\l{fig:Pn-alphalt1-nocond}
\end{figure}

Following the above procedure for parameter values $b=1.75,\alpha=0.5$
that satisfy the conditions (\ref{eq:cond-condensation}) to observe
condensation, Fig. \ref{fig:Pn-alphalt1}(a) shows the
results for the occupancy distribution $\widetilde{p}(n)$ for the full
system with densities
below, at, and above the corresponding critical density $\rho_c$, which is computed numerically from
Eq. (\ref{eq:rhostar}). Consistent with the predictions
of Section \ref{sec:condensation-condition}, we find that a distribution
that decays exponentially for $\rho<\rho_c$ goes over to one decaying as a power law at $\rho=\rho_c$, which at higher densities develops an
additional bump corresponding to the formation of a condensate. The
power-law decay exponent equals $(b+\alpha)$, as predicted in
Eq. (\ref{eq:GC-singlesite-critical-ptilden}). Figure
\ref{fig:Pn-alphalt1}(a) is to be
contrasted with Fig. \ref{fig:Pn-alphalt1-nocond} obtained for the same
disorder realization, system size and $\alpha$, but for a value of $b$ that does not
satisfy the conditions (\ref{eq:cond-condensation}) to observe
condensation; the distribution decays exponentially at all densities.

\begin{figure}[H]
\centering
\includegraphics[width=100mm]{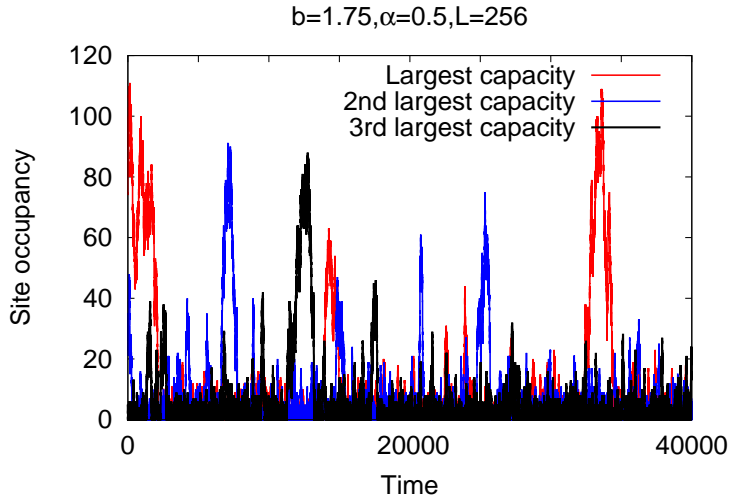}
\caption{For the same disorder realization and other parameters as in
Fig. \ref{fig:Pn-alphalt1}, the figure shows the occupancy at sites with
the largest, the second largest and the third largest capacity, all
plotted together as a function of time for a single dynamical
evolution of the system while starting from the steady state. The data
are obtained by performing Monte Carlo simulations of the dynamics
starting from a steady state configuration.}
\l{fig:timeseries-alphalt1}
\end{figure}

\subsection{Condensate relocation}
In Fig. \ref{fig:Pn-alphalt1}(b), we contrast the
single-site occupancy distribution $p_i(n|l_i)$ for large-capacity sites with
the occupancy distribution $\widetilde{p}(n)$ for the
full system at a density $\rho>\rho_c$. The
single-site distribution is observed to be the same for the site with the largest
capacity (let us denote it by $i_1$) and for the ones with the
second and the third largest capacity (denoted respectively by indices $i_2$ and
$i_3$). To understand such a behavior, we show in Fig.
\ref{fig:timeseries-alphalt1} the results of a MC simulation of the
dynamics for the same set of parameter values and the same disorder
realization as in Fig. \ref{fig:Pn-alphalt1}. The occupancy at sites $i_1,i_2,i_3$ have been
plotted as a function of time for a single dynamical
evolution of the system starting from a steady state configuration. It is
evident from the figure that a
dip in $n_{1_1}$ from a value of $O(L)$ to a value of order $1$ is
followed by a rise within a short time in either
$n_{i_2}$ or $n_{i_3}$ from a value of order $1$ to a value of order
$L$. This implies that the condensate occupies a single site at almost
all times, but does move between certain sites with a relatively small
relocation time. The fact that for a given $L$ and a given disorder realization, there
are only a finite number ${\cal N}(L,\{l_i\}) \sim L^{1-\alpha}$ of sites
that have capacities equal to or larger than $L$ implies that the condensate can relocate
only on this finite subset of sites. A similar relocation dynamics of the condensate on a set of sites whose size grows subextensively with $L$ was observed
in a disordered version of the ZRP studied in \cite{Godreche:2012}, in
which the disorder enters through hop rates.
Such a relocation of the condensate on a subset of sites of
subextensive size may be contrasted with the situation in the
homogeneous ZRP where the condensate can relocate on any of the other
$L-1$ sites \cite{Godreche:2005,Landim:2014}.
In our case, when the condensate has relocated away from 
one of the sites of this subset to another, the occupancy and
fluctuations on the first site
become identical to ones in the background that did not contain the
condensate. This explains why the single-site
distribution for sites $i_1,i_2,i_3$ are the same. A detailed analysis
of the condensate relocation dynamics will be published elsewhere.

\section{Conclusions}
\l{sec:conclusions}
In this paper, we studied a quenched disordered version of
the zero-range process (ZRP), a nonequilibrium system of particles undergoing biased hopping on a one-dimensional
periodic lattice. In the model studied, which we refer to as the random capacity
zero-range process (RC-ZRP), each site has a finite capacity whereby it
can hold only a finite number of particles; we chose the capacities randomly from the Pareto
distribution. We obtained the conditions
for condensate formation in the RC-ZRP, which derive from an interplay of the
capacity distribution with the hop rate. In terms of the power-law
exponents $\alpha>0$ and $b>0$ that characterize respectively the capacity
distribution and the hop rate, we derived explicit conditions for
condensation, namely, $b+\alpha>2$ and $\alpha<1$. Further, we addressed the sample-to-sample variation of the critical density to observe 
condensation, and demonstrated that the corresponding distribution is
either a Gaussian or a L\'{e}vy-stable distribution. 

Let us remark on the possibility of observing condensation in the RC-ZRP for generalizations of the hop rate $u(n)$ that we studied. Consider, e.g., the choice
$u(n)=1+b/n^\sigma$, with $\sigma>0$. For $\sigma<1$, the function $G(l)$
in Eq. (\ref{eq:rhostar}) converges asymptotically to a finite constant
for all values of $b$, yielding a finite $\rho^\star$. As a result, the system
supports condensation for all values of $b$, provided $\alpha<1$, a
condition that derives from the desired scaling of $l_{\rm max}$
with system size $L$. On the other hand, for $\sigma>1$, the function
$G(l)$ diverges asymptotically for all values of $b$, so that $\rho^\star$ is infinite, and consequently, there is no condensate formation in the
system.

We sign off by mentioning a possible follow-up of this work. It would be
of interest to study the RC-ZRP dynamics in the steady state, and investigate the behavior of
time-dependent correlation functions. In this regard, a pertinent issue
is to address if and how quenched disorder manifests itself in the behavior of
the dynamic universality class at criticality and in the dynamics of condensate
relocation, both of which may show significant differences from the
homogeneous ZRP \cite{Gupta:2007}.

\section{Acknowledgments}
SG and MB thank the Galileo Galilei Institute for Theoretical Physics,
 Florence, Italy for hospitality and the INFN for partial support
 during the completion of this work. MB also acknowledges the hospitality of the Rudolf Peierls Centre for Theoretical Physics, University of Oxford, UK. 
SG acknowledges fruitful discussions with
 Martin R. Evans and David Mukamel, and a useful remark on an asymptotic expansion by Pablo Rodriguez-Lopez.

\section{Appendix: Characterizing the site capacities -- The sum and the maximum}
\l{app:Nmax-lmax}
\setcounter{section}{1}
In this appendix, we summarize some features associated with the capacity
distribution (\ref{eq:Pl}) that are relevant to the understanding of the
condensation phenomenon in the RC-ZRP discussed in the main text. 

Let us start with discussing the
behavior of the mean and the variance of the distribution:
they are both finite for $\alpha \ge 2$ and both infinite for $0<\alpha \le 1$. In the
intermediate regime $1 < \alpha <2$, the mean is finite while the
variance is infinite. For values of $\alpha$ in the range $0 <
\alpha \le 2$, the distribution (\ref{eq:Pl}) is L\'{e}vy-stable: a linear combination
of two independently sampled values of $l$ has a distribution identical to $P(l)$, up to location and scale
parameters \cite{Feller1,Feller2,Gnedenko:1955}.  For $0<\alpha<2$, a
L\'{e}vy-stable distribution is characterized by a power-law tail with
exponent $-(1+\alpha)$; the distribution is a
Gaussian for $\alpha=2$.

\begin{figure}[H]
\centering
\includegraphics[width=100mm]{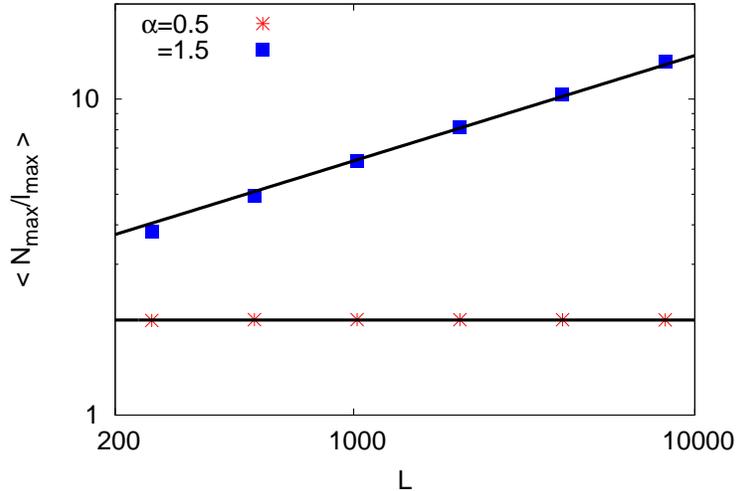}
\caption{Illustrating the validity of the scaling behavior
(\ref{eq:Nmax-lmax-scaling}) of the disorder-averaged ratio $\langle N_{\rm max}/l_{\rm
max} \rangle$ for two values of $\alpha$, one smaller and one larger than
$1$. The data for $\alpha=1.5$ have been scaled down by a
factor of $4$ for convenience of display.}
\l{fig:Nmax-lmax}
\end{figure}

Now, let us discuss the scaling with system size $L$ of the largest
capacity $l_{\rm max}$ and the largest possible number of particles
$N_{\rm max}$ that can be accommodated in the system. For a given $L$
and a given realization $\{l_i\}$ of the disorder, we have $l_{\rm max}\equiv{\rm Max}[l_1,l_2,\ldots,l_L]$ and
$N_{\rm max}\equiv\sum_{i=1}^L l_i$. Since the $l_i$'s are sampled independently from the common distribution
(\ref{eq:Pl}), the probability
distribution of $l_{\rm max}$ is
\bea
{\rm Prob}(l_{\rm max})=LP(l_{\rm max})\Big(\int_1^{l_{\rm
max}}
dl~P(l)\Big)^{L-1}.
\l{eq:lmax-theory}
\eea
In the limit $L \to \infty$, the distribution ${\rm Prob}(l_{\rm max})$
decays for large $l_{\rm max} \gg 1$ as ${\rm
Prob}(l_{\rm max}) \sim \exp(-Ll_{\rm max}^{-\alpha})$, which implies the following scaling of $l_{\rm
max}$ with $L$:
\be
l_{\rm max} \sim L^{1/\alpha},
\l{eq:lmax-scaling}
\ee
valid for all values of $\alpha>0$. 

As to the behavior of $N_{\rm max}$,
for $\alpha >1$, when $P(l)$ has a
finite mean, one may apply the law of large numbers to deduce that
$N_{\rm max}=aL$ in the limit $L \to \infty$, with $a\equiv\int dl~lP(l)$
finite. For $0<\alpha \le 1$,
on the other hand, the mean of $P(l)$ is infinite, the law of large
numbers breaks down, and $N_{\rm max}$ is dominated by contributions from
capacities of order $l_{\rm max}$. Thus, we anticipate $N_{\rm max} \sim
L^{1/\alpha}$, which would imply
\be
\Big\langle \fr{N_{\rm max}}{l_{\rm max}} \Big\rangle \sim \left\{\begin{array}{cc}
          1 & \mbox{for~} 0 < \alpha \le 1, \\
          L^{1-1/\alpha}& \mbox{for~} \alpha > 1.
          \end{array}
          \right.          
\l{eq:Nmax-lmax-scaling}
\ee
Figure \ref{fig:Nmax-lmax} illustrates the validity of the above scaling
for representative values of $\alpha$.
In fact, the full  distribution of the ratio
$N_{\rm max}/l_{\rm max}$ is known (see \cite{Feller2}, page 465), which leads to
\be
\Big\langle \fr{N_{\rm max}}{l_{\rm
max}}\Big\rangle=\fr{\alpha}{1-\alpha};~~0<\alpha<1.
\l{eq:av-Nmax-lmax}
\ee
The above result is confirmed in Fig. \ref{fig:Nmax-lmax}.
Note that Eq. (\ref{eq:Nmax-lmax-scaling}) suggests that there are several
sites other than the site with capacity $l_{\rm max}$ which have capacities of order $L^{1/\alpha}$.


\vspace{1cm}

\begin{thebibliography}{99}
\bibitem{Spitzer:1970}F. Spitzer, Adv. Math. {\bf 5}, 246 (1970).

\bibitem{Evans:2000}M. R. Evans, Braz. J. Phys. {\bf 30}, 42 (2000).

\bibitem{Evans:2005}M. R. Evans and T. Hanney, J. Phys. A: Math. Gen.
{\bf 38}, R195 (2005).

\bibitem{Godreche:2007}C. Godr\`{e}che, in {\it Lecture Notes in
Physics} (Springer-Verlag, Berlin, 2007), Vol. 716; also e-print:arXiv:cond-mat/0604276.

\bibitem{Tripathy:1997}G. Tripathy and M. Barma, Phys. Rev. Lett. {\bf
78}, 3039 (1997).

\bibitem{Tripathy:1998}G. Tripathy and M. Barma, Phys. Rev. E {\bf 58}, 1911 (1998).

\bibitem{Torok:2005}J. T\"{o}r\"{o}k, Physica A {\bf 355}, 374 (2005).

\bibitem{Chowdhury:2000}D. Chowdhury, L. Santen, and A. Schadschneider,
Phys. Rep. {\bf 329}, 199 (2000).

\bibitem{Burda:2002}Z. Burda, D. Johnston, J. Jurkiewicz, M. Kaminski,
M. A. Novak, G. Papp, and I. Zahed, Phys. Rev. E {\bf 65}, 026102
(2002).

\bibitem{Schmittmann:1995} B. Schmittmann and R. K. P. Zia, in {\it Phase Transitions and Critical Phenomena}, edited by C. Domb and J. L. Lebowitz (Academic, London, 1995), Vol. 17.

\bibitem{Stinchcombe:2001}R B Stinchcombe, Adv. Phys. {\bf 50}, 431 (2001).

\bibitem{Schutz:2001}G. M. Sch\"{u}z, in {\it Phase Transitions and Critical Phenomena}, Vol. 19, edited by C. Domb and J. L. Lebowitz (Academic, London, 2001).

\bibitem{Barma:2006}M. Barma, Physica A {\bf 372}, 22 (2006).

\bibitem{Krug:1996}J. Krug and P. A. Ferrari, J. Phys. A: Math. Gen. {\bf 29}, L465 (1996).

\bibitem{Evans:1996}M. R. Evans, Europhys. Lett. {\bf 36}, 13 (1996).

\bibitem{Juhasz:20051}R. Juh\'{a}sz, L. Santen, and F. Igl\'{o}i, Phys. Rev. Lett. {\bf 94}, 010601 (2005).

\bibitem{Juhasz:20052}R. Juh\'{a}sz, L. Santen, and F. Igl\'{o}i, Phys. Rev. E {\bf 72}, 046129 (2005).

\bibitem{Masharian:2012}S. R Masharian and F. H. Jafarpour, Int. J. Mod.
Phys. B {\bf 26}, 1250044 (2012).

\bibitem{Krug:2000}J. Krug, Braz. J. Phys. {\bf 30}, 97 (2000). 

\bibitem{Barma:2002}M. Barma and K. Jain, Pramana - J. Phys. {\bf 58}, 409 (2002).

\bibitem{Jain:2003}K. Jain and M. Barma, Phys. Rev. Lett. {\bf 91}, 135701 (2003).

\bibitem{Harris:2004}R. J. Harris and R. B. Stinchcombe, Phys. Rev. E {\bf 70}, 016108 (2004).

\bibitem{Angel:2004}A. G. Angel, M. R. Evans, and D. Mukamel, J. Stat. Mech. P04001 (2004).

\bibitem{Enaud:2004}C. Enaud and B. Derrida, Europhys. Lett. {\bf 66}, 83 (2004).

\bibitem{Evans:2004}M. R. Evans, T. Hanney, and Y. Kafri, Phys. Rev. E {\bf 70}, 066124 (2004).

\bibitem{Waclaw:2007}B. Waclaw, L. Bogacz, Z. Burda, and W. Janke, Phys. Rev. E {\bf 76}, 046114
(2007).

\bibitem{Grosskinsky:2008}S. Grosskinsky, P. Chleboun, and G. M. Sch\"{u}tz, Phys. Rev. E {\bf 78}, 030101(R) (2008).

\bibitem{Molino:2012}L. C. G. del Molino, P. Chleboun, and S Grosskinsky, J. Phys. A: Math. Theor. {\bf 45}, 205001 (2012).

\bibitem{Godreche:2012}C. Godr\`{e}che and J. M. Luck, J. Stat. Mech. P12013 (2012).

\bibitem{Barma:1993}M. Barma and R. Ramaswamy, in {\it Non-Linearity and
Breakdown in Soft Condensed Matter}, edited by
B. K. Chakrabarti, K. K. Bardhan, and A. Hansen (Springer, Berlin, 1993).

\bibitem{Schutz:1996}G. Sch\"{u}tz, R. Ramaswamy, and M. Barma, J. Phys.
A {\bf 29}, 837 (1996).

\bibitem{Ryabov:2014}A. Ryabov, Phys. Rev. E {\bf 89}, 022115 (2014).

\bibitem{Feller1}W. Feller, {\it An Introduction to Probability Theory
and Its Applications}, (Wiley, New Jersey, 1971), Vol. I.

\bibitem{Feller2}W. Feller, {\it An Introduction to Probability Theory
and Its Applications}, (Wiley, New Jersey, 1971), Vol. II.

\bibitem{Gnedenko:1955}B. V. Gnedenko and A. N. Kolmogorov, {\it Limit
Distributions for Sums of Independent Random Variables},
(Addison-Wesley, Cambridge, Massachussetts, 1954).

\bibitem{Godreche:2005}C. Godr\`{e}che and J. M. Luck, J. Phys. A {\bf
38}, 7215 (2005).

\bibitem{Landim:2014}C. Landim, Commun. Math. Phys. {\bf 330}, 1 (2014).

\bibitem{Gupta:2007}S. Gupta, M. Barma, and S. N. Majumdar, Phys. Rev. E
{\bf 76}, 060101(R) (2007).
\end{thebibliography}
\end{document}